\documentstyle[12pt,aasms4]{article}
\def\etal{{\it et al.\ }}

\def\vs{{\it vs. \,}}

\def\putplot#1#2#3#4#5#6#7{\begin{centering} \leavevmode
\vbox to#2{\rule{0pt}{#2}}
\includegraphics{#1}

\end{centering}}

\begin{document}

\title{Star Formation in Dwarf Galaxies}

\author{Noah  Brosch\altaffilmark{1}}
\affil{Space Telescope Science Institute \\ 3700 San Martin Drive \\
Baltimore MD 21218, U.S.A.}

\and

\author{Ana Heller and Elchanan Almoznino }

\affil{The Wise Observatory and 
the School of Physics and Astronomy \\ Tel Aviv University, Tel Aviv 69978,  
Israel\\
\vspace{10mm}
To be published in {\it the Astrophysical Journal}
}

\altaffiltext{1}{On sabbatical leave from the Wise Observatory and 
the School of Physics and Astronomy,
Raymond and Beverly Sackler Faculty of Exact Sciences,
Tel Aviv University, Tel Aviv 69978, Israel}


\begin{abstract}
We explore mechanisms for the regulation of star formation in dwarf
galaxies. We concentrate primarily on a sample in the Virgo cluster,
which has HI and blue total photometry, for which we collected H$\alpha$ 
data at the Wise Observatory. 
We find that dwarf galaxies do not show the tight correlation of the surface
brightness of H$\alpha$ (a star formation indicator)
 with the HI  surface density, or with the ratio of this
density to a dynamical timescale, as found for large disk
or starburst galaxies. On the other hand, we find the strongest
correlation to be with the average blue surface brightness, indicating 
the presence of a mechanism regulating the star formation by the
older (up to 1 Gyr) stellar population if present, or by the stellar 
population already formed in the present burst.

\end{abstract}


\section{Introduction}
The star formation (SF) is a fundamental process in the evolution of galaxies
and is far from being well understood. The SF is usually characterized by the
initial mass function (IMF) and the total SF rate (SFR),
which depends on many factors such as the density
of the interstellar gas, its morphology, its metallicity, {\it etc.}
According to Larson (1986), four major factors drive star 
formation in galaxies: large scale gravitational instabilities, cloud compression 
by density waves, compression in a rotating galactic disk due to shear
forces, and random cloud collisions. In galaxies with previous stellar
generations additional SF triggers exist, such as shock waves from
stellar winds and supernova explosions.  
In dense environments, such as clusters of galaxies and
compact groups, tidal interactions, collisions with other galaxies,
ISM stripping, and cooling flow accretion probably play some role
in triggering the SF process. The triggering mechanisms were
reviewed recently by Elmegreen (1998).

While ``global''
phenomena, such as the first two SF triggers of Larson (1987), play a large 
part in grand design spirals,
random collisions of interstellar clouds
may provide the best explanation for dwarf galaxies with bursts of SF.
Due to their small size, lack of strong spiral pattern, and sometimes
solid-body rotation ({\it e.g.,} Martimbeau \etal 1994, Blok \&
McGaugh 1997), the star formation
in dwarf galaxies is not triggered by compression due to gravitational 
density waves or by disk shear.
Therefore, understanding SF in dwarf galaxies should be
simpler than in other types of galaxies.

The characterization of the SF processes by a star formation
rate (SFR) controlled by the interstellar gas density as a power law was 
first introduced by Schmidt (1959). The  volume 
density of young stars, $\rho_*$, is related  to the volume 
density of HI gas in the 
Galactic disk as $\rho_*=a \; \rho_{gas}^n$, where $n$ is a constant, 
probably $\approx$2 for spiral galaxies.
In other galaxies the convention is to express the quantities as projected 
densities of stars ($\Sigma_*$) and of gas, as actually observed: 
$\Sigma_*=A \; \Sigma_{gas}^n$. This is usually studied  by 
correlating the surface density of a young star tracer, such as
the H$\alpha$ surface brightness, with the 
gas column density.
 
The H$\alpha$ emission from a galaxy measures its ongoing SFR
(Kennicutt 1983). Gallagher \etal (1984) derived an analytic
relation between the detected H$\alpha$ flux and the present SFR of a 
galaxy; similar relations were derived by Kennicutt \etal (1994).  
The blue  luminosity of a galaxy, on the other hand,  measures its 
past star formation integrated over the last $\sim 10^9$ yrs 
(Gallagher \etal 1984). The newly formed stars, of which the more 
massive produce the Lyman continuum photons which ionize
hydrogen and produce the H$\alpha$ emission, contribute also
to the blue light output of a galaxy. This contribution is minor
in comparison to that from the stars already existing in a
galaxy, unless the SF event is the first in the history of the
galaxy or the star burst is unusually strong. Interestingly, Tresse 
\& Maddox (1998) found recently that the H$\alpha$ luminosity of a 
galaxy correlates with its blue absolute magnitude.

Kennicutt (1998) found that his parametrization of the Schmidt law
fitted well the SF pattern of spiral and IR-selected starburst
galaxies. An alternative to the Schmidt law, proposed
by Silk (1997), fitted equally well. In this variant, the SFR 
per unit area scales with
the ratio of the gas surface density to the local dynamical timescale: $\Sigma_{SFR}\propto\frac{\Sigma_{gas}}{\tau_{dyn}}
\propto\Sigma_{gas}\times\Omega_{gas}$,
 where $\Omega_{gas}$ is the angular rotation speed and the scenario
fits disk configurations.  Kennicutt (1998) adopted
$\Omega_{gas}=\frac{V(R)}{R}$, where V(R) is the rotation
velocity of the gas at a distance R.

Hunter \etal (1998) tested a set of SF predictors on
two small samples of dwarf galaxies, one measured by them and another
derived from de Blok (1997). They found that the ratio of   
$\Sigma_{gas}$  to the critical density for the appearance of ring
instabilities did not correlate with the
star formation, but that the stellar surface brightness did. From this,
they concluded that possibly the stellar energy input provides the
feedback mechanism for star formation.

We concentrate on a sample of late-type dwarf galaxies in the Virgo 
cluster (VC).
The reason for selecting dwarfs was to limit the number of possible trigger
mechanisms of SF; these objects are devoid of large-scale SF triggers, as 
explained above. Having only
VC members limits the sample to a well-defined galaxy background;
in addition, all objects are at $\sim$the same distance and have 
HI information from the 
same source. We tested for correlations between the H$\alpha$ emission
and other observed quantities, in order to investigate mechanisms which
regulate  SF in dwarf galaxies.
The justification to correlate the H$\alpha$ SFR index against  
$\Sigma_{gas}$ is the finding of Kennicutt (1998)
that a Schmidt-type law seems to fit large galaxies. If the SFR
depends on the gas density ratioed to the dynamical
timescale (Silk 1997, Kennicutt 1998), a correlation with 
$\Sigma_{gas}\times\Omega_{gas}$ is expected. Finally, if the SFR 
depends on the local population of blue stars, as found by Hunter \etal 
(1998), then a dependence on the average blue surface brightness is expected.
We also tested the SFR against the ISM gas velocity, 
represented by the width of the HI line profile at 20\% of its peak 
intensity ($\sigma$(HI)), and against a 
combination of it and the surface density of HI, in a manner similar to
that suggested by Silk (1997).
 
\section{The sample}

Our sample consists of 52 late-type dwarf galaxies in the VC 
selected from Binggeli \etal (1985, VCC), with HI measurements from 
Hoffman {\it et al.} (1987, 
1989). The sample was constructed in order to enable the detection
of weak dependencies of the star formation properties on hydrogen content
and surface brightness. We selected two sub-samples
by surface brightness; one represents a high surface brightness (HSB) group
and is either BCD or anything+BCD, and another represents
a low surface brightness (LSB) sample and includes only ImIV or ImV 
galaxies. The morphological classification, which
bins the dwarf galaxies in the HSB or LSB groups, is exclusively from
the VCC. 
In addition, the galaxies are binned by their HI flux integral (FI) from
Hoffman \etal (1987, 1989). The HSB sub-sample
was selected with galaxies of high HI content (FI$>$1500 mJy km s$^{-1}$) or 
with low HI content (0$<$FI$<$600 mJy km s$^{-1}$) and is described in
Almoznino \& Brosch (1998, AB98).
The LSB sub-sample has FI$>$1000 for the high HI sample or 
0$<$FI$<$700 mJy km s$^{-1}$ for the low HI sample (described in Heller 
\etal 1998, HAB98).
The LSB sub-sample is complete, in the sense that it contains all 
objects classified ImIV or ImV in the VCC with m$_B<$17.2 mag. The HSB
sub-sample contains 45\% of the VCC galaxies of this type with 
m$_B<$17.2. Although not complete, it
is representative of this type of object in the VC.

The galaxies were observed at the Wise Observatory (WO) in Mizpe-Ramon
from 1990 to 1997, with CCD imaging through the B, V, R, and I broad bands, 
and narrow H$\alpha$ bandpasses in the rest frame of each galaxy. 
The discussion of all observations and 
their interpretation is the subject of other papers (AB98, HAB98). We 
restrict the discussion here to the analysis
of the integrated H$\alpha$ flux F(H$\alpha$), as it reflects on 
the global process of SF. In particular, we concentrate on 
correlations of this SFR index with other parameters collected 
from the literature.

We  compare our results with other dwarf galaxies for which we
collected published data. We selected Case galaxies from Salzer \etal
(1995, S95) of types HIIH, DHIIH, BCD, MagIrr, and GIrr, as most similar to
our VC sample. We further required that HI observations would exist for the
Case galaxies, and collected seven such objects. As these galaxies do not
have total H$\alpha$ fluxes listed, we estimated those from (a) the total
blue magnitude m$_B$, (b) the listed equivalent width of the H$\beta$ line, and
(c) by assuming $\frac{H\alpha}{H\beta}$=2.9 (Case B, with no extinction). 
A second comparison sample
of eight galaxies originates from Martin (1997), where each object has 
an average H$\alpha$ surface brightness measure in the 1" wide slit. 
FI and $\sigma$(HI)  values
were collected from Huchtmeier \& Richter (1989), while total blue magnitudes
and sizes originate from NED. No corrections for Galactic or
internal extinction were applied to the data. We also assumed that 
the Balmer emission observed spectroscopically is representative of
the entire galaxy.

We prefer to use here distance-independent measures, which are not sensitive
 to the exact location of a galaxy in the VC, to the value of H$_0$,
or to deviations from a smooth Hubble flow, and to stick, as much as
possible, to directly observable quantities. The observables
 F(H$\alpha$), FI, and m$_B$ have, therefore, been normalized 
to the optical area of each galaxy, yielding ``surface brightness'' 
measures per square arcmin. We calculated average blue surface magnitudes
$\Sigma$(B),  average HI flux integrals per unit surface $\Sigma$(HI), 
and average H$\alpha$ surface brightnesses $\Sigma$(H$\alpha$) for all 
objects. The optical area of a galaxy is defined here as
 A=$\frac{\pi D^2}{4R}$,  with D the major axis in arcmin and 
R the axial ratio listed in VCC or estimated from the image of the object
on the Digitized Sky Survey, to yield $\Sigma$(HI). 

We used in some correlations $\sigma$(HI),
and derived $\Omega=\frac{\sigma(HI)}{D}$ as a representative 
gas dynamical property at the outermost optical radius. This definition 
of $\Omega$ is not purely equivalent to that used by Kennicutt (1998), 
but it does not require cosmological assumption 
in its derivation. We caution at this point that $\Sigma$(HI) may
overestimate the surface density of HI in cases where the hydrogen distribution
extends beyond the opical area of an object. Cases where the HI
distribution was 3$\times$ and more larger than the optical size
of a galaxy were reported by Taylor \etal (1995). However, while very 
extended HI distributions do exist, they are not a general characteristic
of dwarf galaxies. Hoffman (private communication) found that only two
of the five Virgo cluster BCDs mapped at Arecibo showed evidence for being
extended. In most cases, the Arecibo beam will cover more than 3$\times$
the optical size of one of our objects, implying that not much HI
could have been missed in the measurements we use here. In absence of 
synthesis or multi-beam mapping of the HI distributions, we selected
to use the coarse measure of $\Sigma$(HI) as defined here, with all
caveats mentioned.

The WO sample ranges over more than two orders of magnitude in $\Sigma$(HI),
over more than three orders of magnitude in $\Sigma$(H$\alpha$), and over
slightly less that two orders of magnitude in $\Sigma$(B). The comparison
sample from Salzer \etal (1995) is more restricted in the range of
H$\alpha$, while the galaxies from Martin (1997) have more
intense H$\alpha$ than the WO objects. In general, galaxies 
from Salzer \etal are $\sim2\times$ more distant than the VC sample,
while objects from Martin are $\sim3\times$ closer than the VC.

\section{Star formation correlations}

We checked first correlations between global parameters of our dwarf
galaxy sample, such as total blue brightness, total HI content, {\it etc.}
In all correlations we considered only detected quantities (no upper limits
were included). We did not find that m$_B$ and the 
HI FIs were correlated in any of the subsamples (for the entire
WO sample the correlation coefficient was 0.57, F=17.4). This
scatterplot is shown in the top left panel of Figure 1. Note that some
degree of correlation would be expected only from the distance effect,
with both m$_B$ and FI being lower for more distant objects.
 
The plot of $\Sigma$(B) vs. $\Sigma$(HI), shown in the top right
panel of Fig. 1,  indicates that 
galaxies with more HI per unit area tend also to have higher blue 
surface brightness, {\it i.e.,} a higher
past-averaged SFR, but this correlation was not very significant. 
We found that log$\Sigma$(H$\alpha$) correlates with the HI line 
width (correlation coefficient 0.61, F=21.3) and show this in the
left middle panel of Fig. 1.
 Dwarf galaxies with brighter blue surface brightness tend also to
have wider HI profiles (correlation coefficient 0.51, F=13.6), as
the right middle panel of Fig. 1 shows. 
$\Sigma$(HI)  correlates also weakly with $\sigma$(HI). This is
illustrated in the lower left panel of Fig. 1.

Kennicutt (1998) showed that in a sample
of large spirals and starburst galaxies
the average H$\alpha$ disk surface brightness correlates well with 
the average molecular and atomic gas surface density.  
Dwarf galaxies have very small quantities of molecular gas ({\it e.g.,}
Gondhalekar \etal 1998), therefore
using here $\Sigma(HI)$ should represent well the total ISM.
This correlation, shown in the lower right panel of Fig. 1, was 
also weak, and the combined WO sample had a correlation
coefficient of only 0.52 (F=13.1)

A better correlation was found for $\Sigma$(H$\alpha$) {\it vs.} the 
``Silk''-type
parameter $\Sigma$(HI)$\Omega$.  Figure 2 shows this for the two VC
samples (HSB=filled diamonds, LSB=squares), 
as well as for the comparison samples from Salzer \etal (1995; 
triangles) and Martin (1997; filled circles).
Note that the two VC samples join up nicely, with the HSB galaxies being
brighter and more H$\alpha$-intense than the LSB objects. 
 The correlation coefficient for
the combined  WO sample is 0.70 (F=34.2) and the slope is
0.93$\pm$0.16.
 The log$\Sigma$(H$\alpha$) correlates even better with the blue
surface magnitude, as Figure 3 shows. For the combined WO sample the 
correlation coefficient 
is 0.77 and the slope is --0.63$\pm$0.09 (F=51.2).

The Salzer \etal (1995) and the Martin (1997) galaxies deviate in both 
plots from the trend set by the VC sample. 
Some of the discrepancy may be the result of our samples being
measured in a uniform and consistent manner,  whereas the plotted 
parameters for the comparison samples were calculated from published 
data and some assumptions (explained above).
The Martin galaxies appear consistently above the location of
the WO galaxies; it is probable that their total H$\alpha$ flux was
over-estimated by assuming that the slit average is representative 
of the entire galaxy. This is confirmed for the three objects in 
common with Marlowe \etal (1997), which have consistently lower
total H$\alpha$ fluxes than adopted by us here.
The Salzer \etal (1995) objects are generally below the WO
objects. They have significant extinction ($<c_{\beta}>\approx0.77$),
which translates into an under-estimate of the H$\alpha$ emission when  
 scaling from the H$\beta$ flux. In addition, the H$\beta$ fluxes
were not corrected for underlying absorption; this also causes an
H$\alpha$ under-estimate. Other reasons for discrepancies may be 
the different distances to the two comparison samples, which influence
$\Omega$ we use here through the angular diameter of a galaxy, used in
the present derivation.

\section{Discussion}

We mentioned above a number of triggers of star formation.
Some, such as shear and two-fluid instabilities, or spiral density waves,
are important mainly in large disk galaxies and thus are not 
relevant for dwarfs.
The sample studied here is comprised of fairly isolated galaxies, although this 
was not a selection criterion. The galaxies are distant enough from other 
objects to
discount recent (few 10$^7$ yrs) interactions as possible star formation triggers.
In these dwarfs the expectation is that the SF may be regulated only by the 
gas density, or by the gas density combined with some factor connected with
the stellar content of the galaxy.
We checked here various correlations of the star formation indicator 
$\Sigma$(H$\alpha$) with global or specific (per unit
area) galaxy parameters. The ``expected'' correlations, observed by 
Kennicutt (1998) to fit well spiral galaxies, were 
found to be much weaker in dwarfs.  The strongest correlation 
was with $\Sigma$(B), while the local ISM dynamic indicator
$\Sigma$(HI)$\Omega$ showed the second strongest correlation.  

The correlation found for the CFRS survey galaxies ($<$z$>\simeq$0.2, 
Tresse \& Maddox 1998),
between the global M$_B$ and log L(H$\alpha$), can be understood if
that survey selected preferentially galaxies of similar sizes in blue and
in hydrogen emission, reducing the problem to a correlation between
area-normalized quantities. These galaxies are much brighter (M$_B\geq$--21
 mag) than the dwarfs discussed here and, being selected on the basis of 
their I-band emission, are probably not representative of the  
``star-forming dwarfs'' class.

Our findings support a scenario whereby the star formation is not controlled 
by the gas volume density, by its surface density, or by the ratio of
the gas surface density to a local dynamical timescale. The strongest
correlation, based on the correlation coefficient and the value of the 
F-statistic, was with the average blue surface magnitude, as found
also by Hunter \etal (1998). There
the question was posed whether this was an effect of the SFR
being $\sim$constant over the last $\sim$1 Gyr. We can definitely rule
out this possibility, as at least one of our objects (VCC 144; Brosch
\etal 1998) seems to exhibit its first SF burst. Many other galaxies,
mainly from the LSB sample, show a number of small HII regions indicating
localized star formation at present. The colors (AB98) are best fitted
by (at least) two stellar populations formed in short bursts, 
 spaced a few 100 Mys to 1 Gyr apart. This indicates that a constant
SF is not a serious possibility for the dwarf galaxies studied here.

\section{Conclusions}
We tested correlations among parameters related to star formation, 
gas and stellar 
content, and internal dynamics on a sample of dwarf galaxies in the Virgo 
cluster. We found that both the Schmidt law and the more recent relation 
derived by Silk (1997) do not fit these galaxies as well as they do 
spirals (Kennicutt 1998). The strongest correlation  of the H$\alpha$
surface brightness, which measures the present star formation
strength, was with the  average blue surface 
brightness, supporting the proposition of Hunter \etal (1998) that
a feedback mechanism must be at work to regulate the present SF 
by the older stellar population.  

\section*{Acknowledgements}
 EA is supported by a special grant from the Ministry of Science and the 
Arts to develop TAUVEX, a UV imaging experiment. AH acknowledges support from the
US-Israel Binational Science Foundation. NB is grateful for continued 
support of the Austrian Friends of Tel Aviv University.  Astronomical 
research at Tel Aviv University
is partly supported by a Center of Excellence award from the Israel Academy of 
Sciences. We acknowledge   Bruno Binggeli
for an updated catalog of the Virgo Cluster and G. Lyle Hoffman for additional
HI information on Virgo galaxies and for comments on a draft of this paper. 
This research has made use of the NASA/IPAC
Extragalactic Database (NED) which  is operated by the Jet Propulsion 
Laboratory, California Institute of Technology, under contract with the 
National Aeronautics and Space Administration. We acknowledge some constructive
remarks, which improved the presentation, by the anonymous referee.

\newpage

\section*{References}
\begin{description}

\item Almoznino, E. \& Brosch, N. 1998, MNRAS, preprint (AB98).

\item Brosch, N., Almoznino, E. \& Hoffman, G.L. 1998, A\&A, 331, 873.

\item Binggeli, B., Sandage, A. \& Tamman, G.A. 1985, A. J., 90, 1681.

\item de Blok, E. 1997, PhD thesis, Rijksuniversiteit Groningen.

\item de Blok, E. \& McGaugh, S.S. 1997, MNRAS, 290, 533.

\item Gallagher, J.S., Hunter, D.A. \& Tutukov, A.V. 1984, ApJ, 284, 544.

\item Gondhalekar, P.M., Johansson, L.E.B., Brosch, N., Glass, I. 
\& Brinks, E. 1998, A\&A, in press.

\item Elmegreen, B.G. 1998, in {\it Origins of Galaxies, Stars, Planets
and Life} (C.E. Woodward, H.A. Thronson, \& M. Shull, eds.), ASP
series, in press.

\item Heller, A., Almoznino, E. \& Brosch, N. 1998, in preparation (HAB98).

\item Hoffman, G.L., Helou, G., Salpeter, E.E., Glosson, J. \& Sandage, A. 
1987, ApJS,  63, 247.

\item Hoffman, G.L., Lewis, B.M., Helou, G., Salpeter, E.E. \& Williams, H.L. 1989 
ApJS,  69, 65.


\item Huchtmeier, W.R. \& Richter, O.-G. 1989, {\it A General Catalog of HI
Observations of Galaxies}, New York: Springer Verlag.


\item Hunter, D.A., Elmegreen, B.G. \& Baker, A.L. 1998, ApJ, 493, 595.


\item Kennicutt, R.C. 1998, ApJ, in press (astro-ph/9712213).

\item Kennicutt, R.C. 1983, ApJ, 272, 54.

\item Kennicutt, R.C., Tamblyn, P. \& Congdon, C.W. 1994, ApJ, 435, 22.

\item Marlowe, A.T., Meurer, G.R. \& Heckman, T.M. 1997, ApJS, 112, 285.

\item Martin, C. 1997, ApJ, 491, 561.

\item Martimbeau, N., Carignian, C. \& Ray, J.-R.  1994, AJ, 107, 543.

\item Salzer, J.J., Moody, J.W., Rosenberg, J.L., Gregory, S.A. \&
Newberry, M.V. 1995, AJ, 110, 920.

\item Schmidt, M. 1959, ApJ, 129, 243.

\item Silk, J. 1997, ApJ, 481, 703.

\item Taylor, C.L., Brinks, E., Grashuis, R.M. \& Skillman, E.D. 1995,
ApJS, 99, 427.

\item Tresse, L. \& Maddox, S.J. 1998, ApJ, in press (astro-ph/9709240).
 
\end{description}

\newpage

\section*{Figure captions}

\begin{description}

\item Figure 1: Scatterplots for various observables measured or 
calculated for the
Virgo cluster sample of dwarf irregular galaxies. The figure shows 
the total HI flux
integral {\it vs.} the total blue magnitude in the top left panel, 
the logarithm
of the HI surface flux integral (HISFI) {\it vs.} the blue surface 
magnitude in the 
top right panel, the logarithm of the H$\alpha$ surface brightness 
($\Sigma$(H$\alpha$)) {\it vs.}
the 20\% width of the HI line in km s$^{-1}$ ($\sigma$(HI))in the middle left 
panel, the blue surface magnitude  {\it vs.} $\sigma$(HI) in the middle right
panel, the HISFI  {\it vs.}  $\sigma$(HI) in the lower left panel, and 
$\Sigma$(H$\alpha$) {\it vs.} $\sigma$(HI) in the lower right panel.
The symbols used are filled diamonds
for the BCD galaxies and squares for the LSBs.

\item Figure 2: Relation between the SFR per unit area 
[log$\Sigma$(H$\alpha$)] and 
log$\Sigma$(HI)$\Omega$, a quantity
proportional to the gas surface density times a global gas-dynamical
measure. The additional symbols with respect to Fig. 1  are filled
circles for objects from Martin (1997) and triangles for galaxies from 
Salzer \etal (1995).
 
\item Figure 3: Relation between the average SFR per unit area
log$\Sigma$(H$\alpha$) and the 
average blue surface magnitude. The symbols are as in Fig. 2.

\end{description}

\newpage

\begin{figure}[htb]
\putplot{/berlin/data1/noah/Papers/DGs/HIvsB.ps}{1in}{-90}{35}{35}{-250}{100}
\putplot{/berlin/data1/noah/Papers/DGs/SHIvsSB.ps}{1in}{-90}{35}{35}{00}{195}
\putplot{/berlin/data1/noah/Papers/DGs/SHavsdv20.ps}{1in}{-90}{35}{35}{-250}{75}
\putplot{/berlin/data1/noah/Papers/DGs/SBvsdv20.ps}{1in}{-90}{35}{35}{0}{170}
\putplot{/berlin/data1/noah/Papers/DGs/SHIvsdv20.ps}{1in}{-90}{35}{35}{-250}{50}
\putplot{/berlin/data1/noah/Papers/DGs/SHavsSHI.ps}{1in}{-90}{35}{35}{0}{145}
\end{figure}

\newpage
   
\begin{figure}[tbh]
\vspace{11cm}
\includegraphics{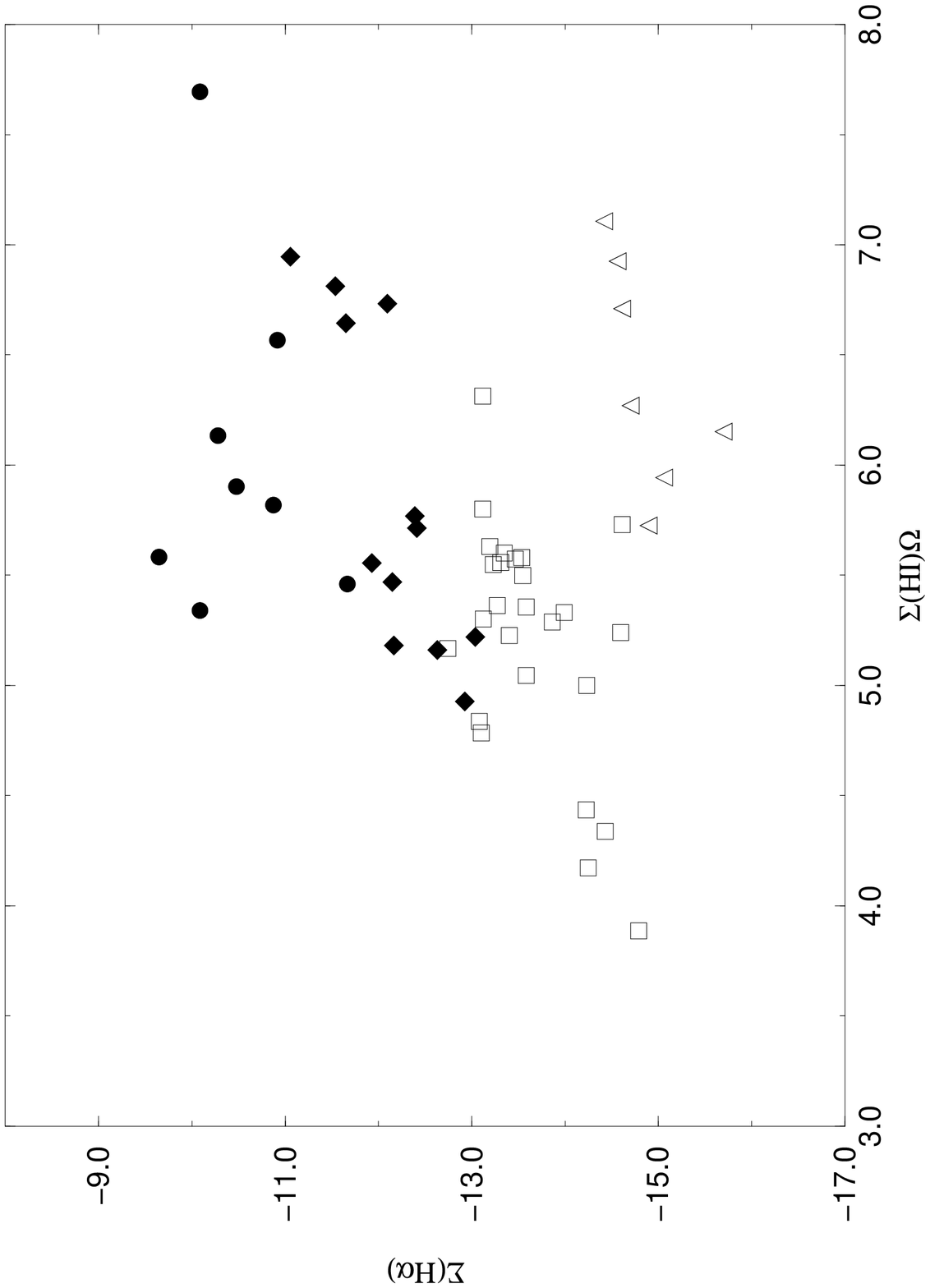}
\end{figure}

\newpage

\begin{figure}[tbh]
\vspace{11cm}
\includegraphics{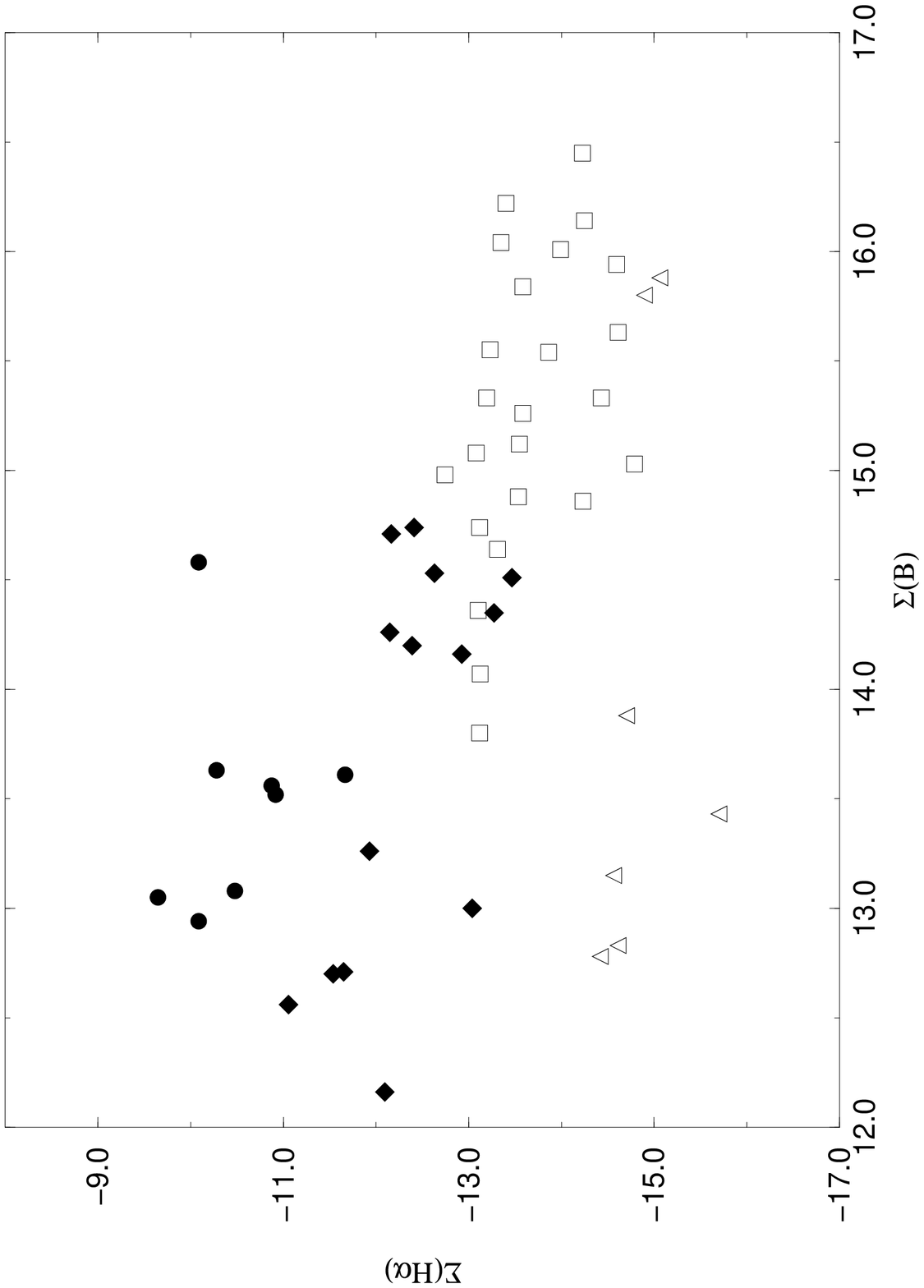}
\end{figure}

\end{document}